\documentclass[12pt]{iopart}

\usepackage{graphicx}
\usepackage{iopams}

\begin{document}

\title{Quantum black holes with charge, colour, and spin at the LHC} 

\author{Douglas M. Gingrich\footnote{Also at TRIUMF, Vancouver, BC V6T
2A3 Canada.}} 

\address{Centre for Particle Physics, Department of Physics,
University of Alberta, Edmonton, AB T6G 2G7 Canada}

\ead{\mailto{gingrich@ualberta.ca}}

\begin{abstract}
In low-scale gravity scenarios, quantum black holes could be produced at
the Large Hadron Collider (LHC) provided the Planck scale is not
higher than a few TeV.
Based on fundamental principles and a few basic assumptions, we have
constructed a model for quantum black hole production and decay in
proton-proton collisions. 
By performing a particle-level simulation at LHC energies, we have
estimated cross sections and branching fractions for final-state
particle topologies that would uniquely identify quantum black holes in
LHC detectors.     
If the Planck scale is about a TeV, even with the most pessimistic
assumptions, the rates for quantum black hole production are estimated
to be above backgrounds, and some of the final-particle states are not
found in Standard Model processes.  
Our results could form the starting point for a detailed investigation
of quantum black holes by the LHC experiments.
\end{abstract}


\section{Introduction\label{sec1}}

Models of low-scale
gravity~\cite{Arkani98,Antoniadis98,Randall99a,Randall99b,Dvali07a,Dvali07b} 
allow for the production of small black holes in particle
collisions~\cite{Argyres98,Banks99,Giddings01a}.  
The available energy must be well above the fundamental Planck scale for
the semiclassical description of black hole production and decay to be
valid.  
Based on current experimental and phenomenological limits on the Planck 
scale~\cite{PDG,Kapner07,ALEPH,DELPHI,L304,LEP,CDF,D0,Hannestad03,Hannestad04,Fairbairn01,Fairbairn02,Kaloper00,Casse04,Anchordoqui01,Anchordoqui03}, 
it is unlikely that semiclassical black holes will be accessible at 
energies produced by the Large Hadron Collider (LHC).   
However, if the Planck scale is low enough, quantum black holes may be
produced in abundance at the
LHC~\cite{Giddings01a,Dimopoulos01,Gingrich08a,Calmet08b}.

In spite of not having a complete theory of quantum gravity, it is
still possible to gain insight into the signatures of quantum black
holes at the LHC based on some fundamental principles and a few
assumptions~\cite{Meade08}.
The current thinking is that a black hole is formed if two partons
from a proton-proton collision at the LHC satisfy the hoop
conjecture~\cite{Thorn02}.
It is natural for the local gauge symmetries of QCD colour and electric
charge to be conserved during the black hole formation and decay
processes~\cite{Calmet08b}.
The black hole thus inherits the colour, charge, and angular momentum
from the parton pair that formed it.
Hence, black holes can be characterized by QCD colour, electric charge,
and angular momentum at the LHC.

Quantum black holes are expected to differ from semiclassical black
holes in a number of ways.
Semiclassical black holes are thought to decay thermally at the Hawking
temperature and be well described by black hole thermodynamics.
In this canonical formalism, the back-reaction on the spacetime during
decay is negligible.  
However, as the black hole mass approaches the Planck scale, the
back-reaction will become significant and the black hole can no longer
be considered in thermal equilibrium with its radiation.
Microcanonical corrections can help improve the decay description until 
quantum mechanical effects take over.
When the Compton wavelength of the black hole becomes larger than the
Schwarzschild radius, quantum effects may impart some particle
characteristics onto the black hole.  
At this point in the decay it is difficult to view the black hole as
still having a well defined temperature and any significant entropy. 
Thus it is unlikely that quantum black holes produced near threshold
will decay thermally~\cite{Alberghi06,Alberghi07,Meade08}.

We will consider black holes produced with mass just above the Planck
scale. 
In this regime, it is unlikely that the decays will be thermal, and we
might expect quantum black hole decays into only a few-particle final
states to dominate.  
The formation and decay will occur over a small region of spacetime,
and the quantum black hole might be viewed as a strongly coupled
resonance or a gravitationally bound state. 
Since the production and decay are short-distance processes, and the
quantum black hole is likely to have colour, the QCD hadronization
process would occur after the decay.

With this simplified picture of quantum black hole production and decay,
we perform a study of the decay signatures that might be expected at the
LHC.  
Our work builds on Ref.~\cite{Meade08,Calmet08b} where these ideas, to
our knowledge, were first discussed.
We explicitly consider only ADD type black holes.
However, there is a range of mass scales for which almost flat
five-dimensional space is an applicable metric for Randall-Sundrum
type-1 black holes.
Some novel signatures that do not often occur in other beyond the
Standard Model physics scenarios will be examined.  
Another unique feature is that, if black holes are produced, these
signatures will occur at huge rates when compared to, even, Standard Model
processes. 

We reminded the reader that there are inevitable limitations in any
model of quantum black holes since this is precisely the regime in which
gravity becomes strongly coupled and the theory is no longer
perturbative.
Thus our seemly predictive results should be viewed as a dimensional
analysis in which it is hope that some of the extrapolations from the
classical domain will carry over to quantum black holes.
With the startup of the LHC, such phenomenological studies are of
interest and of value.  

This paper is structured as follows.
In Sec.~\ref{sec2}, we consider the electric charge, QCD colour, and
spin of black holes produced in proton-proton collisions.
A model for the production of quantum black holes at the LHC is
discussed in Sec.~\ref{sec3} and their decays in Sec.~\ref{sec4}.
In Sec.~\ref{sec5}, we estimate cross sections and discuss the
topologies that might be observed in experiments.

\section{Quantum black hole states\label{sec2}}

Quantum black holes can be classified according to their
$SU(3)_\mathrm{c}$ and $U(1)_\mathrm{em}$ representations.
Since we are considering proton-proton collisions, the allowed particles
forming the black hole are quarks, antiquarks, and gluons.
Nine possible electric charge states can be formed: $\pm 4/3, \pm 1, \pm
2/3, \pm 1/3, 0$.
The 4/3 charge state can only be formed by quark pairs.
The 2/3 charge state can be formed either by an antiquark-antiquark pair
or a quark-gluon combination.
The 1/3 charge state can be formed either by a quark-quark pair or an
antiquark-gluon combination.
The 1 charge state can only be formed by a quark-antiquark pair, and the
0 charge state can be formed by a quark-antiquark pair or a gluon-gluon
pair.
The composition of the black hole negative charge states can be
enumerated similarly. 

The possible colour states of two partons are

\begin{eqnarray} \label{eq1}
3 \otimes \overline{3} & = & 8 \oplus 1\\
3 \otimes 3 & = & 6 \oplus \overline{3}\\
\overline{3} \otimes \overline{3} & = & \overline{6} \oplus 3\\
3 \otimes 8 & = & 3 \oplus \bar{6} \oplus 15\\
\overline{3} \otimes 8 & = & \overline{3} \oplus 6 \oplus \overline{15}\\
8 \otimes 8 & = & 1_S \oplus 8_S \oplus 8_A \oplus 10 \oplus
\overline{10}_A \oplus 27_S\, \label{eq6}. 
\end{eqnarray}

\noindent 
Since black holes form representations of $SU(3)_\mathrm{c}$, they are
predominantly coloured, but can occur as colourless singlets. 
Table~\ref{tab1} lists the possible charge and colour state combinations
that could be produced.   
These states are not unique to quantum black holes but also apply to
semiclassical black holes formed by two-parton collisions.

\begin{table}[heb]
\caption{\label{tab1}Possible quantum black hole states of different
electric charge (superscript) and colour representation (subscript).
For the electric charge states, there are a corresponding 10 states of
opposite electric charge and ``opposite'' colour representation which
are not shown.} 
\begin{indented}
\item[]\begin{tabular}{@{}lllll}
$QBH^{4/3}_{\overline{3}}$  & $QBH^{4/3}_6$ & & &\\
$QBH^1_1$ & $QBH^1_8$ & & &\\
$QBH^{2/3}_3$ & $QBH^{2/3}_{\overline{6}}$ & $QBH^{2/3}_{15}$ & &\\
$QBH^{1/3}_{\overline{3}}$ & $QBH^{1/3}_6$ & $QBH^{1/3}_{\overline{15}}$ & &\\
$QBH^0_1$ & $QBH^0_{8}$ & $QBH^0_{10}$ & $QBH^0_{\overline{10}}$ &
$QBH^0_{27}$\\
\end{tabular}
\end{indented}
\end{table} 

At energies of the fundamental Planck scale $M_D$, the size in spacetime
of the incoming partons and the gravitational radius $r_g$ of the black
hole are both of order $M_D^{-1}$.  
Since the impact parameter is of the same size as the gravitational
radius, the angular momentum of the two-particle system, $J \le M
r_g$, would be of order unity.  
Thus, a possible semiclassical black hole spin-down process is unlikely
to apply to quantum black holes.  
This statement relies on the hoop conjecture of classical gravity being
applicable in the strong gravity regime.

For simplicity, we will consider the initial angular momentum of the
quantum black hole to be due entirely to the spin states of the incoming
partons, and ignore the possibility of an initial small orbital angular
momentum due to an impact parameter.
The procedure outline here can be used to include black holes with
orbital angular momentum in the future, if desired.
When considering spin states, we will assume massless partons.
Since we are ignoring angular momentum (transverse to the helicity axis),
the quantum black hole spin will be parallel to the parton helicity
axis, or the beam axis at the LHC. 
This is quite different from semiclassical black holes with angular
momentum.
In this case, the convention is to add the small amount of spin
angular momentum to the normally larger orbital angular momentum,
which is transverse to the helicity axis~\cite{Frost09}.
The quark-quark states can form spin-0 and spin-1 quantum black holes.
The spin-1 state is three times more likely to form than the spin-0
state. 
The quark-gluon states can form spin-3/2 and spin-1/2 quantum black
holes.
However, because the partons are massless, not all spin combinations are
possible. 
The spin-3/2 state is twice as likely to form as the spin-1/2 state.
The gluon-gluon states can form spin-0, spin-1, and spin-2 quantum black
holes. 
Again because the partons are massless, not all spin combinations are
possible. 
There are no $\pm 1$ states along the spin axis in the spin-2 case.
The relative spin-2, spin-1, and spin-0 probabilities in gluon-gluon
collisions are 7:3:2.
It may well be that charge, colour, and spin are not totally
independent.
The over all wave function of the two parton system may have to be
symmetric or antisymmetric under interchange of two identical quarks or
gluon, respectively.

\section{Production of quantum black holes\label{sec3}}

The cross section for black hole production is not known.
Based on classical arguments and only one available scale, the cross
section is most often taken to be the geometrical cross section $\sigma
\sim \pi r_g^2$, where $r_g$ is the gravitational radius of the
two-particle system (see Ref.~\cite{Gingrich06a} for a review). 
This cross section assumes a radiationless production process in which
all the energy of the two-particle system goes into forming the black
hole.
Various calculations have been performed to predict the amount of
gravitational radiation in the production process, but for
higher-dimensional gravity only the trapped surface approach has yielded
numerical results~\cite{Eardley02,Yoshino02c,Yoshino05a}.
The effect of radiation on the cross section is smallest for low black
hole masses~\cite{Gingrich06a}.
For the quantum black holes that we will consider, the trapped surface
cross section is about 50 time less than the geometrical cross section. 
However, for higher-dimensions and non-zero impact parameter, these
calculations can only be considered as lower bounds on the cross
section. 
In addition, the calculations have ignored parton charge, parton spin,
and parton finite size (see 
Ref.~\cite{Yoshino06a,Gingrich06c,Yoshino07,Kohlprath02} for progress
in these areas). 
The calculations are based on classical arguments, and since it is not
known how applicable these might be in the quantum regime, we will not
consider them except for to realized that the cross sections might be
lower by as much as about two orders of magnitude. 

Arguments have been made for why the geometrical cross section, to
within a small numerical coefficient, should be applicable to quantum
black holes~\cite{Solodukhin02,Hsu03}.
Alternatively, if one embeds a quantum theory of gravity into string
theory, the cross section is reduced due to the quantum size of the
string~\cite{Kohlprath02}.
The energy threshold for quantum black hole production would be
somewhat larger than $M_D$ by a factor that depends on the ratio of
string length to Planck length, or the string coupling in
weakly-coupled string theory. 
The amount of reduction in the cross section will remain unknown until
these fundamental constants have been determined.

We assume the production cross section for quantum black holes can be
extrapolated from the cross section for semiclassical black holes.
It is a challenge for the experiments to measure or set a limit on the
black hole cross section.
This might be one of the few numerical quantities the experiments can
address in an almost model independent way.

The gravitational radius $r_g$ of a quantum black hole of mass $M$
becomes  

\begin{equation}\label{eq7}
r_g = k(D) \frac{1}{M_D} \left( \frac{M}{M_D} \right)^\frac{1}{D-3} , 
\end{equation}

\noindent
where $D$ is the total number of spacetime dimensions, and $k(D)$ is a
numerical coefficient depending only on the number of dimensions and the
definition of the fundamental Planck scale; the PDG definition of the Planck
scale is used in this study:

\begin{equation} \label{eq8}
k(D) = \left( 2^{D-4} \sqrt{\pi}^{D-7}
\frac{\Gamma\left(\frac{D-1}{2}\right)}{D-2} \right)^\frac{1}{D-3} . 
\end{equation}

We now address the question of over what mass range is a black hole a
quantum object?
It is common to take the validity of the semiclassical black hole to be
when the Compton wavelength of the colliding particles lies within the
gravitational radius.
We turn this condition around and consider it to be the upper mass bound
for considering the black hole to be in the quantum regime. 
With this mass restriction, we stay away from the semiclassical
regime, where black hole thermodynamics and thermal decays occur.
This bound also ensure that the initial angular momentum of the quantum
black hole is close to unity.
If we do not take the upper mass requirement into account, the tails of
distributions (particle momentum, for example) might be artificially
altered due to semiclassical black hole decays. 
We consider the lower mass, or threshold, to be when the mass of the
black hole is equal to the inverse of its radius.
Using the PDG convention for the Planck scale gives

\begin{equation} \label{eq9}
\left( \frac{1}{k(D)} \right)^\frac{D-3}{D-2} \lesssim \frac{M}{M_D}
\lesssim \left( \frac{4\pi}{k(D)} \right)^\frac{D-3}{D-2} .
\end{equation}

\noindent
We notice that the minimum mass is below the fundamental Planck scale.
This is an artifact of the definition of the Planck scale.
Using the more intuitive Dimopoulos-Landsberg definition of the Planck
scale gives a minimum
mass always above the fundamental Planck scale.
On the other hand, the accelerator experiments have set limits on
$M_D$ of $M_D \gtrsim 1$~TeV, and hence a quantum back hole would be
require to have mass above only about 500~GeV.
If $M_D$ is about 1~TeV, could it be that the Tevatron has failed to
observe the effects of relatively low-mass black holes?   
Assuming this is not the case, we will take the threshold for quantum
black hole production to be the Planck scale in the PDG definition,
and consider a fixed mass range for all dimensions: $M_\mathrm{min} =
M_D$ and $M_\mathrm{max} = 3M_D$. 
The maximum mass is lower than that indicated by Eq.~(\ref{eq9}), but it
is the lowest common choice where semiclassical approximations are
expected to break down, and thus represents a very conservative maximum
mass for our model.   

Only a fraction of the total centre-of-mass energy $\sqrt{s}$ in a
proton-proton collision is available in the hard scattering process.
We define $s x_a x_b \equiv s x_\mathrm{min} \equiv \hat{s}$, where
$x_a$ and $x_b$ are the fractional energies of the two partons relative
to the proton energies.
The full particle-level cross section $\sigma$ is given by

\begin{equation} \label{eq10}
\sigma(QBH^q_\mathrm{p_1p_2}) = \sum_{a,b} \int^1_{M^2/s} d
x_\mathrm{min} \int^1_{x_\mathrm{min}} \frac{dx}{x} f_a \left(
\frac{x_\mathrm{min}}{x} \right) f_b(x) \pi r_g^2\, , 
\end{equation}

\noindent
where $a$ and $b$ are the parton types in the two protons, and $f_a$ and
$f_b$ are the parton distribution functions (PDFs) for the proton.
The sum is over all the possible quark and gluon pairings that can make
a particular quantum black hole state. 
The parton distributions fall rapidly at high relative energies, and so
the particle-level cross section also falls at high energies.

When describing quantum black hole production, it will be more
informative to specify the two partons p$_1$ and p$_2$ that went into
the formation of the black hole, and drop the colour representation
specifier by summing over the possible colour representations.
This makes sense since the partons hadronize after the quantum black
hole decays.  
The relative probabilities of each colour representation will be
accounted for in the decay branching fractions. 
The resulting 14 different processes can be used to obtained any desired
inclusive result by multiplying the cross section by the branching
fraction and summing the relevant processes.   

\begin{figure}[ht]
\begin{center}
\includegraphics[width=\columnwidth]{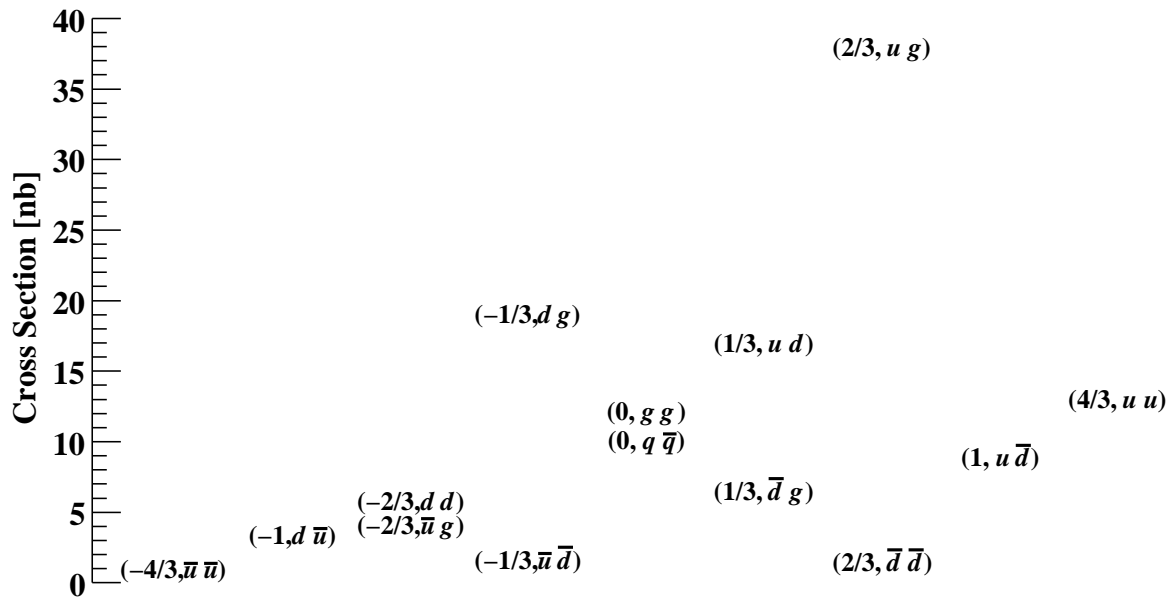}
\caption{\label{fig1}
Cross section spectrum for quantum black holes of different
electric charge and parton progenitors.
$u$ stands for all up-type quarks ($u, c, t$), $d$ stands for all
down-type quarks ($d, s, b$), and similarly for the antiquarks.}
\end{center}
\end{figure}

To obtain numerical results, we have used the parameters shown in
Table~\ref{tab2}.    
You may assume these parameters were used in calculations unless told
otherwise. 
The CTEQ6L1 parton distribution functions~\cite{Pumplin} were used with
a QCD scale of $Q = 1/r_g$.
The cross section spectrum of quantum black hole states is shown in
Fig.~\ref{fig1}. 
It is interesting to note that the highest cross section is given by
$u$-$g$ collisions.
This shows the importance of the gluon contribution at the parton
kinematics and QCD scale that we are using.
As expected, the inclusive cross section, of about 130~nb, is
significantly higher than the semiclassical black hole cross section.
This value should be considered as an upper limit.

\begin{table}[heb]
\caption{\label{tab2}Default parameters used in calculations.}
\begin{indented}
\item[]\begin{tabular}{@{}ccl} \br
Value & Symbol & Description\\ \mr
10     & $D$              & total number of dimensions\\
1~TeV  & $M_D$            & fundamental Planck scale\\
       &                  & (PDG definition)\\
14~TeV & $\sqrt{s}$       & LHC centre-of-mass energy\\
$M_D$  & $M_\mathrm{min}$ & minimum quantum black hole mass\\
$3M_D$ & $M_\mathrm{max}$ & maximum quantum black hole mass\\
\br
\end{tabular}
\end{indented}
\end{table}

The cross section is sensitive to the choice of PDFs and QCD scale.
Table~\ref{tab3} shows the inclusive quantum black hole cross section
for four different choices of PDFs and two different QCD scales.
A difference of about 7\% is observed for the different PDFs.
Our calculations use $Q = 1/r_g$ for the QCD scale.
If $Q = M$ is used, the cross sections are about 13\% lower. 

\begin{table}[ht]
\caption{\label{tab3}Inclusive quantum black hole cross section in
nanobarns for different parton density functions and different QCD
scales.} 
\begin{indented}
\item[]\begin{tabular}{@{}ccccc} \br
QCD   & \multicolumn{4}{c}{Parton Density Functions}\\\cline{2-5}
Scale & CTEQ6L1 & CTEQ5L & CTEQ5M & MRST98\\ \mr
$Q = 1/r_g$ & 132 & 136 & 159 & 150\\
$Q = M$     & 115 & 119 & 139 & 132\\
\br
\end{tabular}
\end{indented}
\end{table}

\begin{figure}[ht]
\begin{center}
\includegraphics[width=\columnwidth]{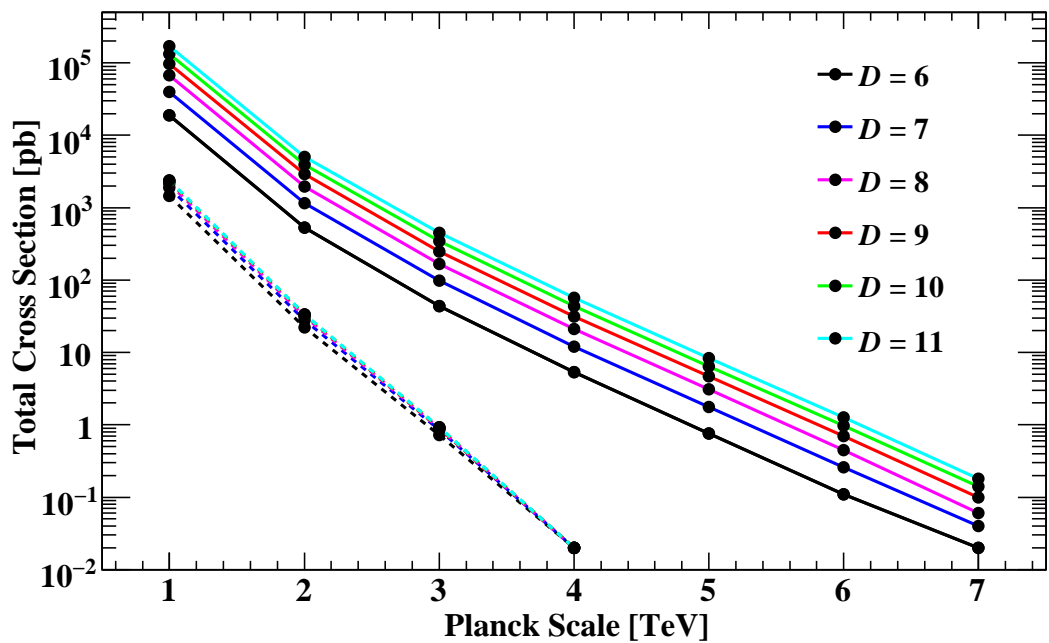}
\caption{\label{fig2}
Total proton-proton cross section for different number of
dimensions $D$ and different fundamental Planck scales.
The solid lines are totally inelastic cross sections and the dashed
lines are trapped surface cross sections.
$M_\mathrm{max} = \mathrm{min}(3M_D,\sqrt{s})$ has been used.}
\end{center}
\end{figure}

Figures~\ref{fig2} shows how the total inclusive cross section changes
with different number of dimensions and different values for the Planck
scale.  
If the LHC detectors can collect a few hundred pb$^{-1}$ of data in the
first years, they will be able to produce significant numbers of
quantum black holes for Planck scales as high as about 5~TeV, if the
geometrical cross section is valid.  
Also shown in Fig.~\ref{fig2} are the trapped surface lower-bound
cross sections.
These cross sections are about 10 to $10^4$ lower than the geometrical 
cross section over the range $1 < M_D < 4$~TeV.
In this case, we see that there is very little dependence on the
number of dimensions.
In calculating the trapped surface cross section, the quantum black
hole mass has been limited to the range of $M_D < M < 3 M_D$.
That is, if the mass before gravitonal radiation is above $3M_D$ it is 
not included in the total cross section calculation.

\begin{figure}[ht]
\begin{center}
\includegraphics[width=\columnwidth]{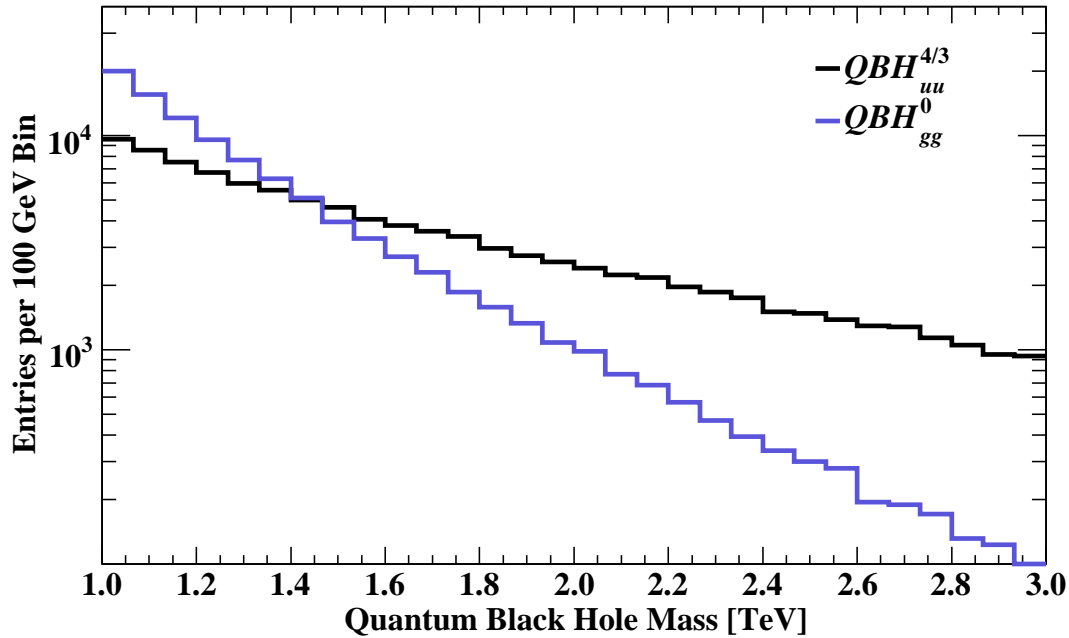}\
\caption{\label{fig3}
Quantum black hole mass distribution for two different
quantum black hole states.} 
\end{center}
\end{figure}

Because of the falling product of parton distribution functions with
parton-parton centre-of-mass energy $\sqrt{\hat{s}}$, the most probable
value for the quantum black hole mass is $M_D$.
At this value, the gravitational radius, and hence cross section, is
independent of the quantum black hole mass.
Thus the shape of the cross section at the lowest possible masses is
almost independent of the parton cross section and is determined
predominantly by the parton distribution functions.
Since each quantum black hole state is made from unique valance quarks,
sea quarks, or gluons, the product of parton density functions can be
very different.  
Figure~\ref{fig3} shows the mass distribution for two rather different 
quantum black hole states. 
Although gluons contribute the most at low masses, the $u$-quarks
(valance quarks) contribute the most above masses of about $1.5 M_D$.
Over the mass region $1 < M < 3$~TeV, the $uu\to QBH^{4/3}$ differential
cross section falls as a power law in the mass with exponent $-2.1$,
while the $gg\to QBH^0$ cross section falls as a power law in the mass
with exponent $-4.6$.  

\begin{figure}[ht]
\begin{center}
\includegraphics[width=\columnwidth]{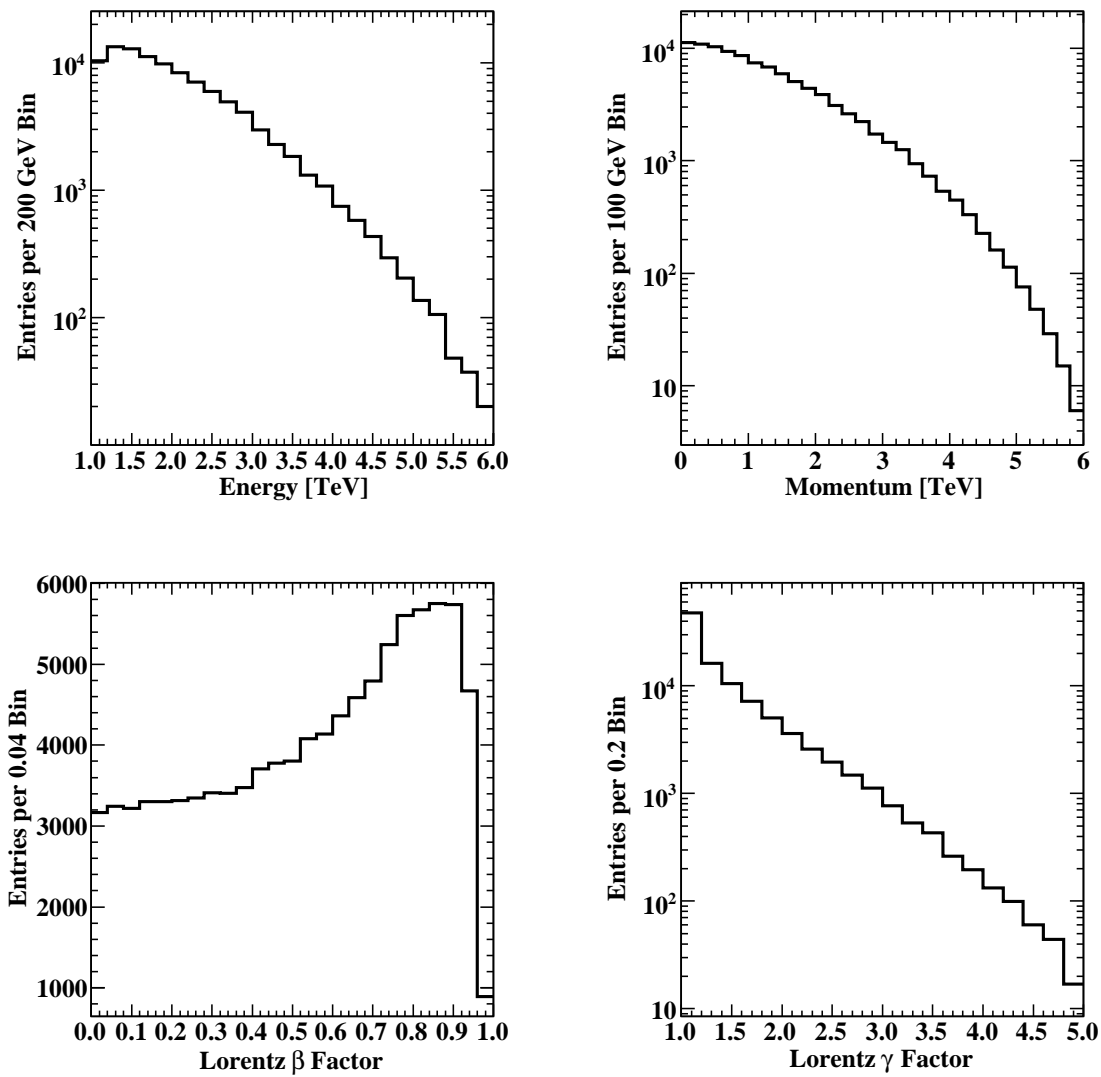}
\caption{\label{fig4}
Distributions of some kinematic variables of quantum black
holes.}
\end{center}
\end{figure}

Since quantum black holes are predominantly produced near threshold, their
kinematics are different from the kinematics of semiclassical black holes.
Figure~\ref{fig4} shows the kinematics of quantum black holes.
The mean mass is about $1.5 M_D$, the mean energy is about $2 M_D$, and the 
mean momentum is about $0.9 M_D$.
The distribution of values of the Lorentz variables $\beta$ and
$\gamma$ can be easily explained. 
They depend only on the $x_a$ and $x_b$ values of the two partons forming
the black hole, or equivalently on the $\hat{s}$ and $s$ values.
$\beta$ and $\gamma$ are given by
 
\begin{equation} \label{eq11}
\beta = \frac{|x_a-x_b|}{x_a+x_b} 
\quad \mathrm{and} \quad
\gamma = \frac{x_a + x_b}{2\sqrt{x_a x_b}}\, .
\end{equation}

\noindent
For symmetric collisions ($x_a\approx x_b$), 
$\beta \approx 0$ and $\gamma \approx 1$.
For highly asymmetric collisions ($x_1 \approx 1$  and $x_2 \approx
x_\mathrm{min}$),

\begin{equation} \label{eq12}
\beta \approx \frac{|s - \hat{s}|}{s + \hat{s}} \approx 0.99, \quad 
\gamma \approx \frac{s + \hat{s}}{2\sqrt{s\hat{s}}} \approx 
\frac{\sqrt{s}}{2\sqrt{\hat{s}}} \approx
\frac{E_\mathrm{beam}}{2~\mathrm{TeV}} \approx 7\, .
\end{equation}

\noindent
Thus, based on Fig.~\ref{fig4} we can see that the collisions are more
asymmetric than symmetric. 

\section{Quantum black hole decay\label{sec4}}

Up until now, we have visualized the hard scattering process as
occurring through a quantum black hole resonance; in analogy to a
particle resonance. 
It is informative to estimate the width of this possible resonance.
For a well defined resonance, we will use the criteria that $\tau \gg
1/M$, where $M$ is the initial quantum black hole mass and $\tau$ is its
lifetime.
It is assumed that the quantum black hole decays entirely without
leaving a remnant.  
When estimating the lifetime, it is usual to assume Hawking emission of
massless particles during the entire decay process. 
One can start with the power spectrum for black hole emission or go
directly to the generalized  Stefan-Boltzmann equation and relate the
Hawking temperature to the Schwarzschild radius.
The total power emitted will be equal to the rate of decrease of the
black hole mass.
The resulting differential equation is then integrated over the entire
lifetime of the black hole and the result is given by

\begin{equation} \label{eq13}
\tau = \frac{1}{CM} \left( \frac{M}{M_D} \right)^\frac{2(D-2)}{D-3} ,
\end{equation}

\noindent
where the form of $C$ depends on how the emissivities of the different
fields are normalized. 
In any case, $C$ involves a sum over the particle degrees of freedom
for each spin weighted by the integrated power emitted into each spin
field. 
The power factors are greybody-modified thermal power spectra proposed
by Hawking~\cite{Hawking75}.

Since we are considering non-thermal quantum black hole decays, the
applicability of Eq.~(\ref{eq13}) is questionable.
However, we will consider it as an indicative estimate of the quantum
black hole lifetime.
Since quantum black holes are produced with mass close to the Planck
scale, we expect $\tau M \gtrsim 1/C$.
Thus our requirement for a resonance translates into $C \ll 1$ or large
mass (at which point the black hole is semiclassical).
The coefficient $C$ increases with the number of particle degrees of
freedom, the value of the greybody factors, and the number of
dimensions. 
If  greybody factors are ignored, $C \approx 2$ for $D = 10$ and does not
drop below unity until $D < 8$.
If we use the greybody factors calculated according to
Ref.~\cite{Harris03b} and only consider emission on the brane,  
$C \approx 5$ for $D = 10$, and $C$ is still greater than two for seven
dimensions. 
If we include the emission of gravitons into the bulk using the results
of Ref.~\cite{Cardoso06a}, the number of degrees of freedom increases,
particularly for higher dimensions, and $C$ becomes even larger.
Thus, unless we push the limits of the quantum black hole mass into the
semiclassical regime or restrict our considerations to low number of
extra dimensions (those likely to be excluded by experiments), a
quantum black hole probably can not be viewed as a particle resonance.
Some authors have reached similar conclusions~\cite{Harris05b,Meade08}.
The concept of a quantum black hole state still has the useful purpose
of labeling the possible interactions.

In the decays of quantum black holes, one expects the number of
particles in the final state to be small. 
Using arguments similar to those used to estimate the lifetime, the
average number of particles emitted from the black hole during Hawking
evaporation can be estimated to be

\begin{equation}\label{eq14}
\langle N \rangle = \rho S \sim \left( \frac{M}{M_D}
\right)^\frac{D-2}{D-3} ,  
\end{equation}

\noindent
where $S$ is the initial entropy.
The form of $\rho$ depends on how the emissivities of the different
fields are normalized. 
If greybody factors are ignored, the average multiplicity is between
1.0 to 1.3 for 6 to 10 dimensions.
If greybody factors are included and emission is restricted to the
brane, there is no significant change in the average multiplicities.
Allowing graviton emission into the bulk gives the numbers shown in
the second column of Table~\ref{tab4}.
The mean multiplicity increases with the number of dimensions to a
maximum for nine dimensions and then decreases. 
This is due to the interplay between the entropy definition (due to
the Planck scale definition) and the ratio of the sum of fluxes to sum
of powers in the coefficient $\rho$ in Eq.~(\ref{eq14}).

\begin{table}[ht]
\caption{\label{tab4}Mean number of emissions $\langle N\rangle$ and
probability of each number of particles in the final state versus the
number of dimensions $D$.}
\begin{indented}
\item[]\begin{tabular}{@{}ccccccc} \br
$D$ & $\langle N\rangle$ & \multicolumn{5}{c}{Number of Particles}\\
\cline{3-7}
& & 2 & 3 & 4 & 5 & 6\\ \mr
 5 & 0.6 & 0.74 & 0.21 & 0.04 & 0.01 & 0.00\\
 6 & 0.9 & 0.61 & 0.28 & 0.09 & 0.02 & 0.00\\
 7 & 1.1 & 0.54 & 0.30 & 0.11 & 0.03 & 0.01\\
 8 & 1.2 & 0.51 & 0.31 & 0.13 & 0.04 & 0.01\\
 9 & 1.3 & 0.50 & 0.31 & 0.13 & 0.04 & 0.01\\
10 & 1.2 & 0.51 & 0.31 & 0.13 & 0.04 & 0.01\\
11 & 1.1 & 0.54 & 0.30 & 0.11 & 0.03 & 0.01\\
\br
\end{tabular}
\end{indented}
\end{table}

Fluctuations about the mean multiplicity can be described using a
Poisson distribution~\cite{Bekenstein05,Anchordoqui03}.
The Poisson distribution can also be used to estimate the relative
probabilities of two-particle, three-particle, four-particle, etc.\
final states. 
When calculating the relative probabilities, we have removed the case
of zero particle emission and renormalized the Poisson distributions.
We consider single-particle emission to represent two-particle decay
in that the remaining black hole is considered to be the second
particle.  
Using these concepts, the probabilities for different number of
particles in the final state are shown in Table~\ref{tab4}.
For $D=7$ to 11 there is little dependence on the number of dimensions.
Approximately 50\% of the decays are two-particle, while three-particle
and four-particle decays are not insignificant. 
The multiplicities depend on the definition of the Planck scale.
For the Dimopoulos-Landsberg definition and the case of $D = 10$, the
mean multiplicity is 0.4 and the probability of a two-particle decay is
about 80\%.
The two-particle state is enhanced because, by using a different
definition of the Planck scale, we are effectively considering a lower
threshold for quantum black hole production. 
In what follows, we will only consider two-particle final states.
Clearly, an extended analysis should include higher multiplicity 
decays. 

\begin{figure}[ht]
\begin{center}
\includegraphics[width=\columnwidth]{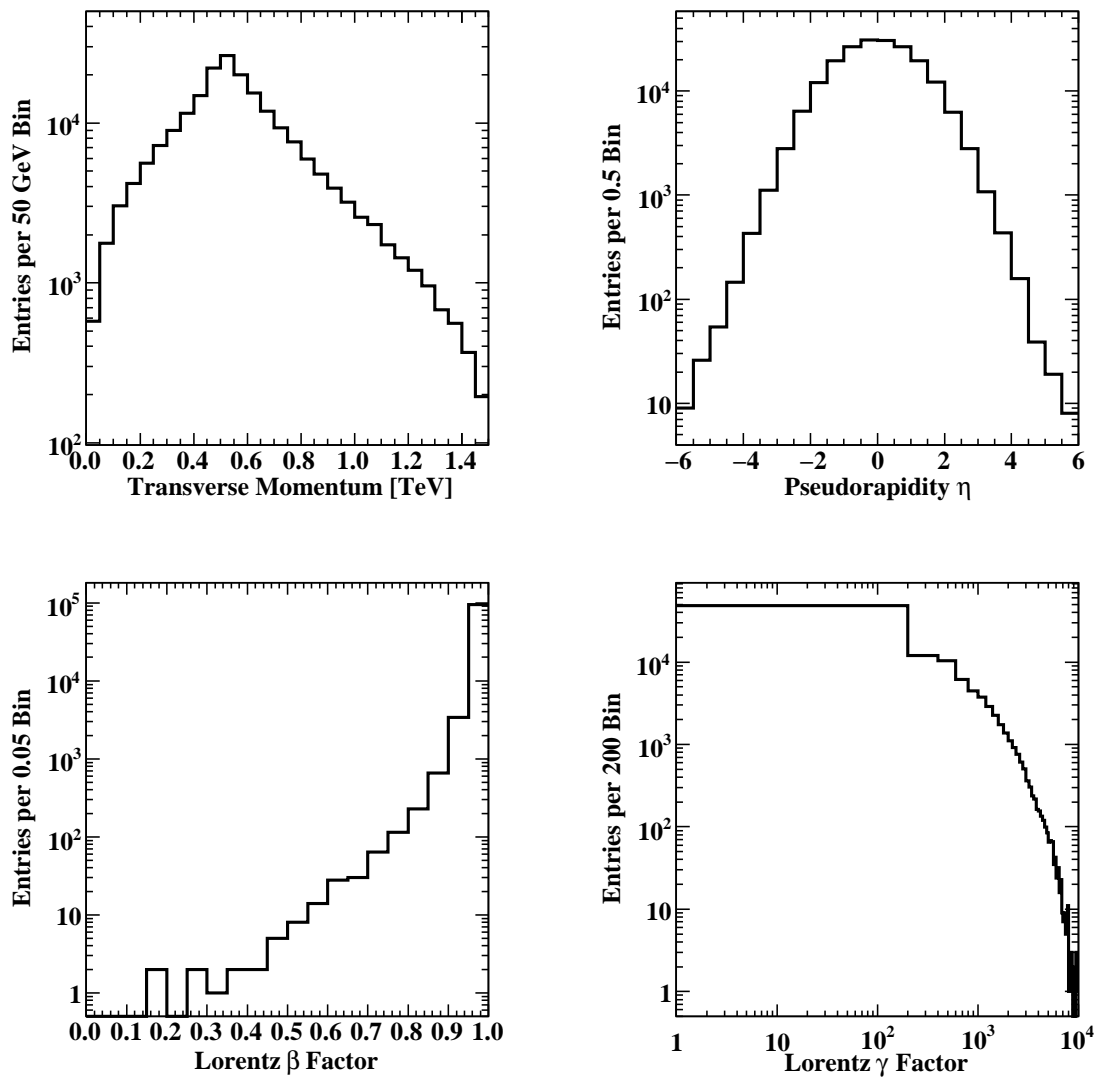}
\caption{\label{fig5}
Distributions of some kinematic variables for decay particles
from quantum black holes.
Massless particles, electrons, muons, $u$-quarks, and $d$-quarks have
not been included in the Lorentz $\beta$ and $\gamma$ distributions.}
\end{center}
\end{figure}

For the decay kinematics, we have used two-particle phase space decays.
The decay particle kinematics are shown in Fig.~\ref{fig5}.
In the Lorentz $\beta$ and $\gamma$ distributions, we have excluded
zero-mass particles and low-mass particles: electrons, muons,
$u$-quarks, and $d$-quarks. 
Since we are using two-particle decays, the distributions are
representative of all particle types, except for the distributions of
the Lorentz variables $\beta$ and $\gamma$.
The most probable particle energy, momentum, or transverse momentum is
about $M_D/2$.
Most particles will remain in a detector and are highly relativistic.  

In quantum black hole decays, we consider the colour and charge to be
conserved, but make no similar assumptions about global charges like
baryon or lepton number.  
We consider gravity to be totally democratic and couple to all flavours
of quarks, leptons, and gauge bosons equally.
(See Ref.~\cite{Dvali08b,Dvali08c,Dvali09} for a discussion of
non-democracy in a model with many particle species.)
Although perhaps debatable, we assume Lorentz invariance holds, and
only allow decays that conserve total angular momentum.

Experimental bounds on the effects of higher-dimension operators might
limit the parameter space over which we can consider decays violating
global charges. 
The limits on flavour-changing lepton decays and proton decay should be
taken into consideration.
Unfortunately, it is not clear what the relationship is between the
Planck scale and the scale occurring in higher-dimension operators.
Arguments have been made for why the experimental constraints are
serious but do not necessarily rule out the allowed parameter
space that we will consider~\cite{Meade08,Calmet08b}.
It is also possible that the global symmetries may be gauged.

Baryon and lepton violating processes may also be mediated by virtual
black holes and thus affect higher order diagrams~\cite{Grumiller04}.
The probability of proton decay mediated by a virtual black hole is very
small for regular high-scale gravity in four
dimensions~\cite{Zeldovich76}.  
However, in low-scale gravity, the baryon number violating processes
through formation of an intermediate virtual black hole can become
sizable.
Various methods have been proposed to suppress these baryon and lepton
violating processes to agree with experimental
limits~\cite{Arkani00,Bambi07}. 

Quantum black holes have colour, and size of about $M_D^{-1} \lesssim 1~ 
\mathrm{TeV}^{-1}$. 
The decay products can be partons in coloured representations, which
can travel a distance of about the QCD scale of a Fermi before they have
to hadronize to form colour singlets.
Thus, we only allow parton decay final states that can have the colour
representation of the quantum black hole, or equivalently, the colour
representation of the initial parton progenitors.  
Since it would be difficult to identify the quantum black hole colour
in experiments, we will consider the different singlet and octet
representations (see Eq.~(\ref{eq1}) and (\ref{eq6})) to be the same when
counting the number of degrees of freedom. 

The particle content of low-scale gravity is uncertain.
In fact, a large number of dark sector species may reveal themselves in
the strong gravity regime~\cite{Dvali07a,Dvali07b,Dvali09}.
A Higgs boson might be discovered and there is likely to be a graviton.
In addition, how the neutrino sector couples to gravity is an
interesting topic. 
Now that at least one neutrino has mass, it is possible that gravity
will couple to neutrinos of both left- and right-handedness equally.
In addition, whether neutrinos are Dirac or Majorana particles will
change the number of available final states.
However, the biggest factor effecting the final state particles is if
the global symmetries of the Standard Model remain good symmetries at
the strong gravity scale. 
In presenting our results, we most often consider two extreme particle
content models.
Our base model allows global symmetries to be violated, includes a
Higgs and graviton, and assumes the neutrinos are Majorana particles of
both handedness. 
Our comparison model is the currently observed Standard Model: global
symmetries are conserved, there is no Higgs or graviton, and the
neutrinos are left-handed Dirac particles.
When considering the different particle content models it is important
to realized that coupling per degree of freedom in Hawking radiation
does not apply in the quantum gravity cases.
For quantum gravity, we are simply assuming the coupling to all quark and
lepton flavours are equal.

We determined the branching fractions for each quantum black hole state as
follows. 
For each state, we wrote down all the two-particle final states that
could conserve colour, charge, and angular momentum. 
Some decay states can not conserve all the angular momentum modes and
were thus weighted accordingly.
For each decay mode, we formed a product of weights given by the number
of members in the colour representation, times the number of flavour
combinations, times the spin degrees of freedom.
The decay branching fractions are shown Table~\ref{tab5}. 

\begin{table}[ht]
\caption{\label{tab5}Quantum black hole decay branching fractions. 
BR is the branching fraction, C means global symmetries are conserved
and V means they may be violated.}
\begin{indented}
\item[]\begin{tabular}{@{}lclcclclcc} \br
State & & Decay & \multicolumn{2}{c}{BR (\%)} &
State & & Decay & \multicolumn{2}{c}{BR (\%)}\\ \mr
& & & C & V & & & & C & V\\ \mr
$QBH^{4/3}_{uu}$             & $\to$ & $uu$             & 100 & 67  &
$QBH^0_{q\bar{q}}$ & $\to$ & $u\bar{u}$     & 41.5& 36.5\\
                             &       & $\bar{d}\ell^+$  &     & 33  &
                 &       & $d\bar{d}$       & 41.5& 36.5\\\cline{1-5}
$QBH^1_{u\bar{d}}$           & $\to$ & $u\bar{d}$       &  86 & 79.8&
                 &       & $gZ$             & 4.1 & 3.6\\
                             &       & $\nu\ell^+$      &   3 & 8.9 &
                 &       & $gg$             & 4.1 & 3.6\\
                             &       & $W^+g$           &   9 & 7.9 &
                 &       & $g\gamma$        & 4.1 & 3.6\\
                             &       & $W^+Z$           &   1 & 1.0 &
                 &       & $\ell^+\ell^-$   & 1.5 & 4.1\\
                             &       & $W^+\gamma$      &   1 & 1.0 &
                 &       & $\nu\nu$         & 1.2 & 2.7\\
                             &       & $W^+H$           &     & 0.7 &
                 &       & $W^+W^-$         & 0.5 & 0.5\\
                             &       & $W^+G$           &     & 0.7 &
                 &       & $\gamma\gamma$   & 0.5 & 0.5\\\cline{1-5}
$QBH^{2/3}_{ug}$             & $\to$ & $ug$             &  73 & 66.7&
                 &       & $ZZ$             & 0.5 & 0.5\\
                             &       & $dW^+$           &   9 & 8.3 &
                 &       & $\gamma Z$       & 0.5 & 0.5\\
                             &       & $u\gamma$        &   9 & 8.3 & 
                 &       & $g H$            &     & 2.7\\
                             &       & $uZ$             &   9 & 8.3 &
                 &       & $\gamma H$       &     & 0.3\\
                             &       & $uH$             &     & 2.8 &
                 &       & $Z H$            &     & 0.5\\
                             &       & $uG$             &     & 5.6 &
                 &       & $H H$            &     & 0.1\\\cline{1-5}
$QBH^{2/3}_{\bar{d}\bar{d}}$ & $\to$ & $\bar{d}\bar{d}$ & 100 & 50  &
                 &       & $g G$            &     & 2.7\\
                             &       & $u\nu$           &     & 25  &
                 &       & $\gamma G$       &     & 0.3\\
                             &       & $d\ell^+$        &     & 25  &
                 &       & $Z G$            &     & 0.3\\\cline{1-5}
$QBH^{1/3}_{\bar{d}g}$       & $\to$ & $\bar{d}g$       &  73 & 66.7&
                 &       & $G G$            &     & 0.5\\\cline{6-10}
                             &       & $\bar{u}W^+$     &   9 & 8.3 &
$QBH^0_{gg}$     & $\to$ & $u\bar{u}$       & 27.8& 27.1\\
                             &       & $\bar{d}\gamma$  &   9 & 8.3 &
                 &       & $d\bar{d}$       & 27.8& 27.1\\
                             &       & $\bar{d}Z$       &   9 & 8.3 &
                 &       & $gg$             & 27.9& 27.1\\
                             &       & $\bar{d}H$       &     & 2.8 &
                 &       & $gZ$             & 7.0 & 6.8\\
                             &       & $\bar{d}G$       &     & 5.6 &
                 &       & $g\gamma$        & 7.0 & 6.8\\\cline{1-5}
$QBH^{1/3}_{ud}$             & $\to$ & $ud$             & 100 & 60  &
                 &       & $\ell^+\ell^-$   & 0.5 & 1.6\\
                             &       & $\bar{d}\nu$     &     & 20  &
                 &       & $\nu\nu$         & 0.3 & 1.1\\
                             &       & $\bar{u}\ell^+$  &     & 20  &
                 &       & $W^+W^-$         & 0.4 & 0.4\\\cline{1-5}
&&&&&                 &       & $\gamma\gamma$   &  0.4 &  0.4\\
&&&&&                 &       & $ZZ$             &  0.4 &  0.4\\
&&&&&                 &       & $\gamma Z$       &  0.4 &  0.4\\
&&&&&                 &       & $Z H$            &      &  0.1\\
&&&&&                 &       & $H H$            &      &  0.1\\
&&&&&                 &       & $H G$            &      &  0.2\\
&&&&&                 &       & $G G$            &      &  0.4\\
\br
\end{tabular}
\end{indented}
\end{table}  

We reminded the reader that the quantitative results presented in
Table~\ref{tab5} and the following section are of an illustrative
nature. 
They are based on many assumptions and extrapolations from the
semi-classical regime.

\section{Results\label{sec5}}

Using the previously developed model, we have simulated the production
and decay of quantum black hole events, with each state weighted by
its cross section, using a new Monte Carlo event generator described
in Ref.~\cite{Gingrich09a}.
To out knowledge no such study has yet been performed.
The generators Charybdis~\cite{Frost09} and BlackMax~\cite{Dai07} both
include some facility to generate quantum black holes.
BlackMax implements the suggestions of Meade and Randall
verbatim~\cite{Meade08}.
The decay species are chosen according to thermal greybody-modified
distributions, which we have argued might not be applicable to quantum
black hole decays.   
Charybdis allows the user to set parameters so that effectively a
quantum black hole decaying to two-particle final states is generated,
but does not take colour explicitly into account when determining the
two-particle final state.
In both cases, the generators do not allow for the ability to generate
a particular quantum black hole. 
They can thus be made to act as inclusive quantum black hole
generators. 
In addition, these generators do not allow baryon number violating
processes that can be subsequently hadronized.
The new generator can be used to select the specific quantum black hole
state to generate, includes all possible decay modes properly
weighted by conserving $SU(3)_c$ and $U(1)_\mathrm{em}$, but can violate
baryon number.
   
\begin{figure}[ht]
\begin{center}
\includegraphics[width=\columnwidth]{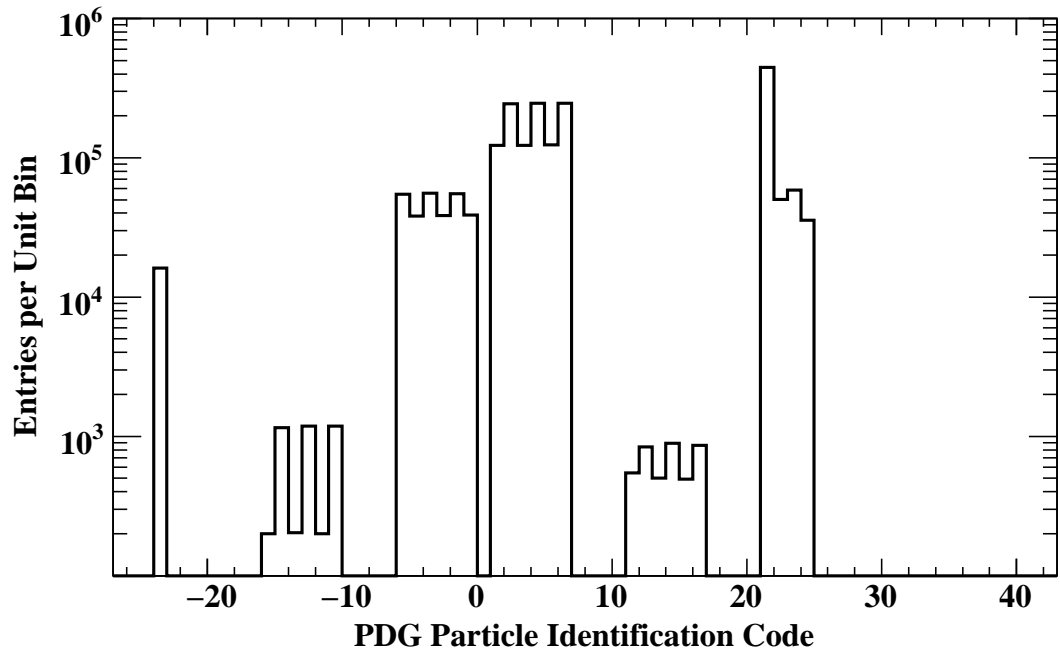}
\caption{\label{fig6}
Relative occurrence of decay particle types according to
their PDG identification code.
Observed Standard Model particle content with global symmetries
conserved.}
\end{center}
\end{figure}

\begin{figure}[ht]
\begin{center}
\includegraphics[width=\columnwidth]{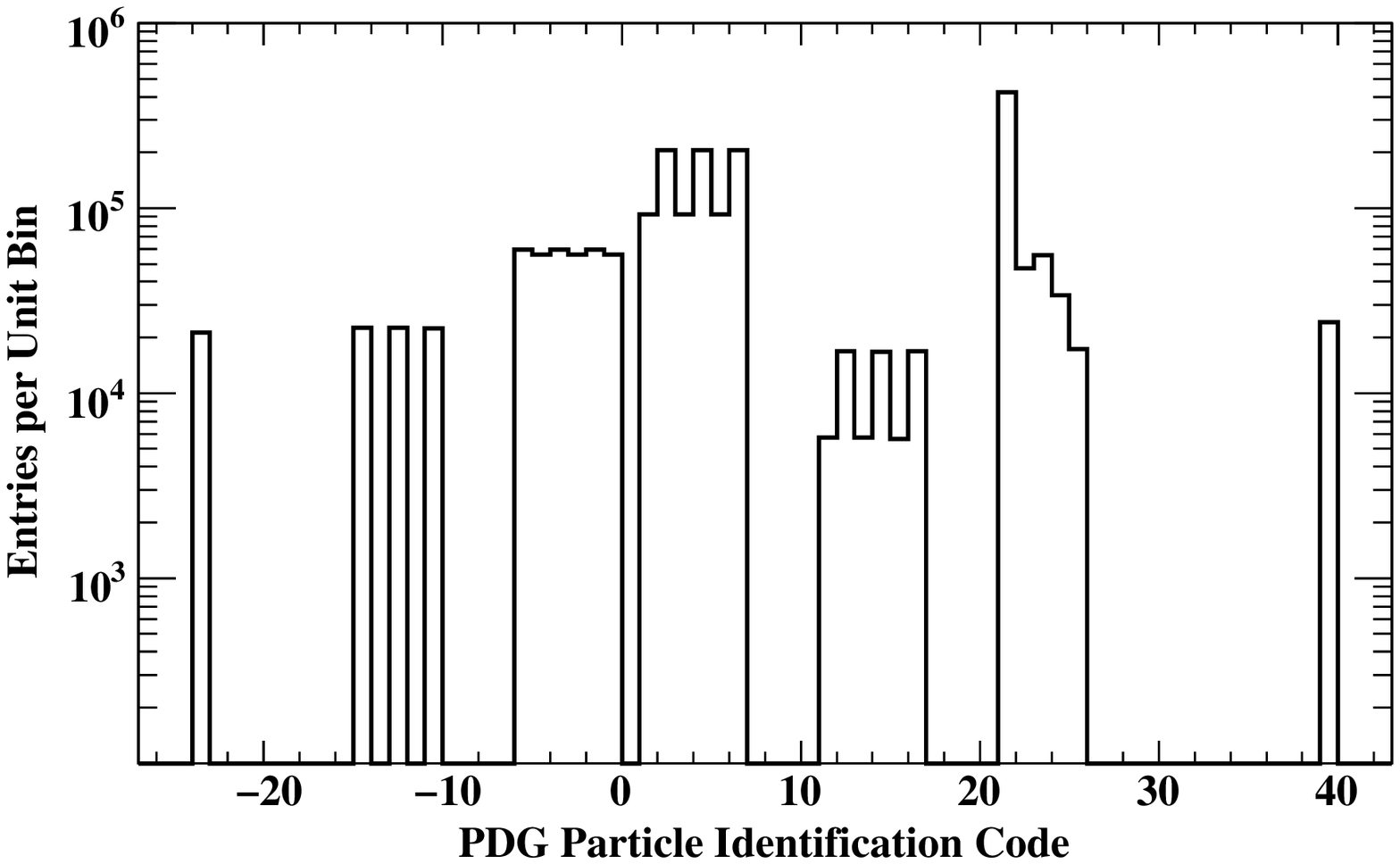}
\caption{\label{fig7}
Relative occurrence of decay particle types according to
their PDG identification code.
Global symmetries may be violated, neutrinos are chiral and Majorana,
and a Higgs and graviton are allowed.}
\end{center}
\end{figure}

The particle PDG identifier codes for the two decay particles in the
hard scattering are shown in Fig.~\ref{fig6}. 
We notice the expected dominance of quarks and gluons, and the charge
and baryon asymmetry from having two protons collide, rather than
a proton-antiproton collision. 
Figure~\ref{fig6} is for the observed Standard Model particle content
and can look significantly different depending on the particle
content model. 
If neutrinos are chiral, the number of charged leptons and neutrinos are
approximately equal.
If global symmetries can be violated, the number of charged antileptons
and antineutrinos increases, while the number of antiquarks decreases. 
Including the graviton is significant, while the Higgs boson has little
effect. 
Figure~\ref{fig7} show the frequency of particle identification codes
when including all these effects.
We notice a significant increase in leptons and near equality between
the different antiquark flavours.

We have studied the multiplicities of various particle signatures in a
detector: jets, electrons, muons, photons, and missing energy. 
For this calculation, we allowed the $t$, $W$, $Z$, and $H$ to decay.  
We called all quarks, gluons, and tau particles jets.
On average per event there are 2.4 jets, 0.05 electrons, 0.05 muons,
0.05 photons, and 0.2 particles (neutrinos and gravitons) that give
missing energy. 
There is very little variation due to changing the particle content
model.
If global symmetries can be violated, the average number of electrons,
muons, and particles giving missing energy per event are marginally
higher. 
This is because of the domination of jets in both cases.

It is useful to know the average number of jets in events and how the
decays of the $t$-quark, $W$, and $Z$ effect these numbers:
99\% of the events have at least one jet, 87\% of the event at least
two jets, and 28\% of the events at least three jets.
If global symmetries are violated, the number of events with two or more
jets decreases by about 10\%.
Parton hadronization, detector effects, and jet-finding algorithms
will cause the number of jets per event to be higher.

To study the topologies of the events, we do not allow the final-state
particles to decay.
The percentage occurrence of each topology is shown in
Table~\ref{tab6}. 
As expected, di-jet events dominate.
Also significant are mono-jet, jet plus $t$-quark, jet plus
gauge-boson, and jet plus lepton topologies. 
Noteworthy is the electron plus muon topology at the 0.08\% level.
Such a signal is not usually produced in other beyond the Standard
Model physics processes.
The cross sections are also shown in Table~\ref{tab6}, along with
predictions of the Standard Model when using
PYTHIA~8~\cite{pythia8,pythia64}.
For the Standard Model processes, we have restricted the phase-space
by requiring the invariant mass to be between 1~TeV and 3~TeV, and the
transverse momentum in the rest frame of the hard scattering process
to be between 500~GeV and 1.5~TeV. 
In most cases, the Standard Model cross sections are more than three
orders of magnitude lower.
So even if the quantum black hole cross sections are at the lower
bound given by the trapped surface calculation, the number of events
produced in each decay topology should be greater than the Standard
Model background. 
 
\begin{table}[ht]
\caption{\label{tab6}Percentage occurrence and cross section for each
decay topology. 
Columns two and three are for quantum black holes that conserve global
symmetries and columns four and five are for quantum black holes 
that may violate global symmetries.
The last column is for Standard Model QCD and electroweak processes.}
\begin{indented}
\item[]\begin{tabular}{@{}lccccc} \br
Topology & \multicolumn{2}{c}{$B, L$ Conserved} &
\multicolumn{2}{c}{$B, L$ Violated} & Standard Model\\\cline{2-6}  
& (\%) & $\sigma$ (pb) & (\%) & $\sigma$ (pb) & $\sigma$ (pb)\\ \mr  
di-jets        & 64    & $8.5\times 10^4$ & 54 & $7.2\times 10^4$ & 630\\
jet + $t$      & 19    & $2.6\times 10^4$ & 14   & $1.9\times 10^4$ & 1.4\\
jet + $Z$      & 4.3   & $5.7\times 10^3$ & 4.0  & $5.2\times 10^3$ & 1.2\\ 
jet + $W$      & 4.3   & $5.6\times 10^3$ & 4.4  & $5.9\times 10^3$ & 2.8\\ 
jet + $\gamma$ & 4.3   & $5.7\times 10^3$ & 3.9  & $5.2\times 10^3$ &
$8.9\times 10^{-1}$\\  
$t\bar{t}$     & 2.8   & $3.6\times 10^3$ & 2.0  & $2.7\times 10^3$ & $1.7\times 10^{-1}$\\
mono-jets      & 0.072 &  95 & 7.5   & $1.0\times 10^4$ &\\
jet + $\mu$    &       &     & 3.5   & $4.7\times 10^3$ &\\
jet + $e$      &       &     & 3.5   & $4.6\times 10^3$ &\\
jet + $H$      &       &     & 1.3   & $1.8\times 10^3$ &\\
no energy      & 0.098 & 130 & 0.29  & 380 &\\
mono-$\mu$     & 0.070 &  92 & 0.24  & 320 &\\
mono-$e$       & 0.070 &  92 & 0.24  & 310 &\\
di-$Z$         & 0.056 &  74 & 0.052 &  69 & $5.2\times 10^{-3}$\\ 
di-$W$         & 0.056 &  75 & 0.052 &  69 & $2.7\times 10^{-2}$\\ 
$Z$ + $W$      & 0.068 &  89 & 0.077 & 100 & $1.0\times 10^{-2}$\\ 
di-$\gamma$    & 0.053 &  70 & 0.054 &  71 & $2.2\times 10^{-3}$\\ 
$\gamma$-$Z$   & 0.056 &  75 & 0.050 &  66 & $4.1\times 10^{-3}$\\
$\gamma$-$W$   & 0.067 &  89 & 0.086 & 110 & $2.8\times 10^{-3}$\\
di-$\mu$       & 0.042 &  56 & 0.037 &  49 &\\
di-$e$         & 0.045 &  59 & 0.039 &  51 &\\
$e$ + $\mu$    &       &     & 0.079 & 100 &\\
$W$ + $H$      &       &     & 0.058 &  77 & $2.6\times 10^{-3}$\\
$Z$ + $H$      &       &     & 0.035 &  47 & $1.3\times 10^{-3}$\\
mono-$W$       &       &     & 0.058 &  77 & $1.5\times 10^{-1}$\\
mono-$H$       &       &     & 0.011 &  15 & $1.5\times 10^{-4}$\\
mono-$Z$       &       &     & 0.022 &  30 & $6.3\times 10^{-2}$\\
mono-$\gamma$  &       &     & 0.028 &  37 & $2.9\times 10^{-2}$\\
$H$ + $H$      &       &     & 0.014 &  18 &\\
$H$-$\gamma$   &       &     & 0.023 &  52 &\\
\br
\end{tabular}
\end{indented}
\end{table}

The previous analysis used a Planck scale of 1~TeV, which is a hard
lower bound for $D > 6$.
If one considers how Kaluza-Klein gravitons would affect supernovae
cooling and neutron stars, the lower bounds on the Planck scale are
higher~\cite{Hannestad03,Hannestad04}.
However, it should be realized that the bounds provided by
astrophysical arguments contain significant uncertainties, such as the
compactification moduli.
For $D < 8$, the lower bound on the fundamental Planck scale is too
high to allow quantum black holes to be produced by the LHC.
For higher dimensions, the bounds on the Planck scale are less stringent.
For $D = 8$, $M_D > 4$~TeV and for $D > 8$, $M_D > 1.4$~TeV.
For $D > 8$, the lower bound on the Planck scale is set by the absence
of black holes in neutrino cosmic ray showers~\cite{Anchordoqui03}.
However, Auger has yet to observe a single neutrino-induced shower.
Ref.~\cite{Anchordoqui03} is based on the ratio of vertical and
quasi-horizontal neutrino showers.
The vertical showers are used to normalize the product of the neutrino
flux and their interaction cross section.
The fact that Auger should have seen a few vertical showers by now,
but have not seen any yet, relaxes the limits on $M_D$.
In addition, the neutrino cosmic ray bounds can be evaded in a model of
split fermions~\cite{Stojkovic06}.
Using the results of Fig.~\ref{fig2}, we see that for $D = 8$, $\sigma
\sim 20$~pb and for $D > 8$, $\sigma \gtrsim 2\times 10^4$~pb.
Thus, most decay signatures in Table~\ref{tab6} would be observable for
$D > 8$, while probably none would be observable for $D < 8$.

We comment on the $D = 8$ case a little more by giving some order
of magnitude estimates.
We assume the ratio of the number of observed events to the
experimental acceptance is $10^3$.
For di-jet events, the over all acceptance might be high, but some
background can be anticipated and hence on the order of 100 events
will need to be observed.
While in the case of leptons, the backgrounds might be lower, but the
acceptance will probably be lower since more stringent requirements
might be needed to reduce lepton fake rates, and ensure the leptons
are not accompanied by additional jets.
Thus about 80~pb$^{-1}$ of data would be be required to see an anomaly
in the Standard Model di-jet events, and about 1~fb$^{-1}$ to see one
in a jet plus lepton signal.  
About 50~fb$^{-1}$ of data would be needed for a definitive statement to
be made in the electron plus muon channel.

\section{Summary}

Since many assumptions have been made in developing our toy model, we 
conclude by summarizing them.
The model is based on assuming local charges (colour, electric
charge, and spin) are conserved in the strong gravity regime of black
hole production and decay near the Planck scale.
However, we have not taken any initial orbital angular momentum of the
black holes into account, but argued that it would be small.  
We have only examined two-body final states and motivated this choice by   
extrapolating semiclassical results to the Planck scale.
We have required Lorentz invariance and have not considered a black hole 
remnant -- both arbitrary, but working choices.
Conservation of global charges like baryon and lepton number has not
been required.
Predictions based on global charge conservation and non-conservation
have been presented.

Experiments at the LHC are currently preparing to search for the
effects of low-scale gravity below the Planck scale by searching for
gravitational scattering and resonances, and above the Planck scale by
searching for semiclassical black holes.
However, there has been little guidance in the literature as to the
phenomenology of low-scale gravity at the Planck scale. 
This is because of the quantum and non-perturbative nature of gravity in
this strong-gravity regime. 
To help guide experimentalists, we have considered black hole production
and decay near the Planck scale.
Based on fundamental principles and some assumptions, we have built a
model to estimate cross sections and decay topologies, albeit aware of
its limitations. 
Although di-jet decays denominate, they are unlikely to be the only
decays with a significant rate.
The jet plus gauge-boson and jet plus $t$-quark topologies account for 
about 30\% of the decays. 
In models where global symmetries need not be conserved, jets plus
leptons and mono-jets are also significant decays.
If the Planck scale is low enough so that the inclusive cross section
is of the order of 100~nb, then even the decays with low branching
fractions will be readily observable.
Signatures such as opposite sign electron plus muon, with very little
else in the detector, will be hard to explain by conventional physics.
Observing anomalous rates in several of the decay channels could help
determine the nature of the new physics and lead to the discovery of
quantum gravity. 

\ack

I would like to thank Xavier Calmet and De-Chang Dai for helpful e-mail
discussions. 
This work was supported in part by the Natural Sciences and Engineering
Research Council of Canada.



\section*{References}
\begin{thebibliography}{00}

\bibitem{Arkani98}
Arkani-Hamed N, Dimopoulos S and Dvali G 
1998 
{\em Phys. Lett. B}
{\bf 429}
263

\bibitem{Antoniadis98}
Antoniadis I, Arkani-Hamed N, Dimopoulos S and Dvali G 
1998
{\em Phys. Lett. B}
{\bf 436}
257

\bibitem{Randall99a}
Randall L and Sundrum R
1999
{\em Phys. Rev. Lett.}
{\bf 83}
3379

\bibitem{Randall99b}
Randall L and Sundrum R
1999
{\em Phys. Rev. Lett.}
{\bf 83}
4690

\bibitem{Dvali07a}
Dvali G 
2007
{\em Preprint} 0706.2050

\bibitem{Dvali07b}
Dvali G and Redi M 
2008 
{\em Phys. Rev. D}
{\bf 77}
045027

\bibitem{Argyres98}
Argyres P C, Dimopoulos S and March-Russell J
1998
{\em Phys. Lett. B}
{\bf 441}
96

\bibitem{Banks99}
Banks T and Fischler W
1999
{\em Preprint} hep-th/9906038

\bibitem{Giddings01a}
Giddings S B and Thomas S
2002
{\em Phys. Rev. D}
{\bf 65}
056010

\bibitem{PDG}
Yao W-M \textit{et al} (Particle Data Group)
2006
{\em J. Phys. G: Nucl. Part. Phys.}
{\bf 33}
1

\bibitem{Kapner07}
Kapner D J, Cook T S, Adelberger E G, Gundlach J H,
Heckel B R, Hoyle C D and Swanson H E 
2007
{\em Phys. Rev. Lett.}
{\bf 98}
021101

\bibitem{ALEPH}
Heister A  \textit{et al} (ALEPH Collaboration)
2003
{\em Eur. Phys. J. C}
{\bf 28}
1

\bibitem{DELPHI}
Abdallah J \textit{et al} (DELPHI Collaboration)
2005
{\em Eur. Phys. J. C}
{\bf 38}
395

\bibitem{L304}
Achard P \textit{et al} (L3 Collaboration)
2004
{\em Phys. Lett. B}
{\bf 587}
16

\bibitem{LEP}
LEP Exotica Working Group, ALEPH, DELPHI, L3 and OPAL Collaborations
2004
Combination of {LEP} results on direct searches for large extra
dimensions
{\em CERN Note LEP Exotica WG} 2004-03.

\bibitem{CDF}
Abulenica A \textit{et al} (CDF Collaboration)
2006
{\em Phys. Rev. Lett.}
{\bf 97}
171802

\bibitem{D0}
Abazov V M \textit{et al} (D{\O} Collaboration)
2008
{\em Phys. Rev. Lett.}
{\bf 101}
011601

\bibitem{Hannestad03}
Hannestad S and Raffelt G G 
2003 
{\em Phys. Rev. D}
{\bf 67}
125008

\bibitem{Hannestad04}
Hannestad S and Raffelt G G 
2004
{\em Phys. Rev. D}
{\bf 69}
029901

\bibitem{Fairbairn01}
Fairbairn M
2001
{\em Phys. Lett. B}
{\bf 508}
335

\bibitem{Fairbairn02}
Fairbairn M and Griffiths L M 
2002
{\em J. High Energy Physics}
{\bf 02}
024

\bibitem{Kaloper00}
Kaloper N, March-Russell J, Starkman G D and Trodden M
2000
{\em Phys Rev. Lett.}
{\bf 85}
928

\bibitem{Casse04}
Cass{\'e} M, Paul J, Bertone G and Sigl G
2004
{\em Phys. Rev. Lett.}
{\bf 92}
111102

\bibitem{Anchordoqui01}
Anchordoqui L A, Feng J L, Goldberg H and Shapere A D 
2002
{\em Phys. Rev. D}
{\bf 65}
124027

\bibitem{Anchordoqui03}
Anchordoqui L A, Feng J L, Goldberg H and Shapere A D 
2003
{\em Phys. Rev. D}
{\bf 68}
104025

\bibitem{Dimopoulos01}
Dimopoulos S and Landsberg G
2001
{\em Phys. Rev. Lett.}
{\bf 87}
161602

\bibitem{Gingrich08a}
Gingrich D M and Martell K
(2008)
{\em Phys. Rev. D}
{\bf 78}
115009

\bibitem{Calmet08b}
Calmet X, Gong W and Hsu S D H 
2008
{\em Phys. Lett. B}
{\bf 668}
20

\bibitem{Meade08}
Meade P and Randall L
2008
{\em J. High Energy Physics}
{\bf 05}
003

\bibitem{Thorn02}
Thorn K S
2002 
Nonspherical gravitational collapse: A short review
in J. R. Klauder, Magic Without Magic, San Fancisco.

\bibitem{Alberghi06}
Alberghi G L, Casadio R, Galli D, Gregori D, Tronconi A and Vagnoni V
2006
arXiv:hep-ph/0601243v1

\bibitem{Alberghi07}
Alberghi G L, Casadio R and Tronconi A
2007,
{\em J. Phys. G: Nucl. Part. Phys.}
{\bf 34}
767

\bibitem{Frost09}
Frost J A, Gaunt J R, Sampaio M O P, Casals M, Dolan S R, Parker M A and
Webber B R 
2009
{\em J. High Energy Physics}
{\bf 10}
014

\bibitem{Gingrich06a}
Gingrich D M 
2006
{\em Int. J. Mod. Phys. A}
{\bf 21}
6653

\bibitem{Eardley02}
Eardley D M and Giddings S B 
2002
{\em Phys.\ Rev.\ D}
{\bf 66}
044011

\bibitem{Yoshino02c}
Yoshino H and Nambu Y
2003 
{\em Phys. Rev. D}
{\bf 67}
024009

\bibitem{Yoshino05a}
Yoshino H and Rychkov V S 
2005
{\em Phys.\ Rev.\ D}
{\bf 71}
104028

\bibitem{Yoshino06a}
Yoshino H and Mann R B 
2006 
{\em Phys.\ Rev.\ D}
{\bf 74}
044003

\bibitem{Gingrich06c}
Gingrich D M 
2007
{\em J. High Energy Phys.}
{\bf 02}
098

\bibitem{Yoshino07}
Yoshino H, Zelnikov A and Frolov V P 
2007
{\em Phys. Rev. D}
{\bf 75}
124005

\bibitem{Kohlprath02}
Kohlprath E and Veneziano G
2002
{\em J. High Energy Phys.}
{\bf 06}
057

\bibitem{Solodukhin02}
Solodukhin S N 
2002
{\em Phys.\ Lett.\ B}
{\bf 533}
153

\bibitem{Hsu03}
Hsu S D H 
2003
{\em Phys. Lett. B}
{\bf 555}
92

\bibitem{Pumplin}
Pumplin J, Stump D, Huston J, Lai H L, Nadolsky P and Tung W K  
2002
{\em J. High Energy Phys.}
{\bf 07}
012

\bibitem{Hawking75}
Hawking S W 
1975
{\em Commun. math. Phys.}
{\bf 43}
199

\bibitem{Harris03b}
Harris C M and Kanti R
2003
{\em J. High Energy Phys.}
{\bf 10}
014

\bibitem{Cardoso06a}
Cardoso V, Cavagli{\`a} M and Gualtieri L
2006
{\em J. High Energy Phys.}
{\bf 02}
021

\bibitem{Harris05b}
Harris C M 
Physics Beyond the Standard Model: Exotic Leptons and Black Holes
at Future Colliders, 
PhD thesis, University of Cambridge, December, 2003,
arXiv:hep-ph/0502005v1.

\bibitem{Bekenstein05}
Bekenstein J D and Mukhanov V F 
1995
{\em Phys. Lett. B}
{\bf 360}
7

\bibitem{Dvali08b}
Dvali G
{\em Preprint} 0806.3801

\bibitem{Dvali08c}
Dvali G and Pujol{\`a}s O
2009
{\em Phys. Rev. D}
{\bf 79}
064032

\bibitem{Dvali09}
Dvali G and Redi M
2009
{\em Phys. Rev. D}
{\bf 80}
055001

\bibitem{Grumiller04}
Grumiller D
2004
{\em Int. J. Mod. Phys. D}
{\bf 13}
1973

\bibitem{Zeldovich76}
Zeldovich Ya B 
1976
{\em Phys. Lett. A}
{\bf 59}
254

\bibitem{Arkani00}
Arkani-Hamed N and Schmaltz M 
2000
{\em Phys. Rev. D}
{\bf 61}
033005

\bibitem{Bambi07}
Bambi C, Dolgov A D and Freese K  
2007
{\em Nucl. Phys. B}
{\bf 763}
91

\bibitem{Gingrich09a}
Gingrich D M 
{\em Preprint} 0911.5370.

\bibitem{Dai07}
Dai, D-C, Starkman G, Stojkovic D, Issever C, Rizvi E and Tseng J
2008
{\em Phys. Rev. D}
{\bf 77}
076007

\bibitem{pythia8}
Sj{\"o}strand T, Mrenna S and Skands P
2008
{\em Comput. Phys. Commun.}
{\bf 178}
852

\bibitem{pythia64}
Sj{\"o}strand T, Mrenna S and Skands P
2006
{\em J. High Energy Phys.}
{\bf 0605}
026

\bibitem{Stojkovic06}
Stojkovi D, Starkman G D and Dai D-C 
2006
{\em Phys. Rev. Lett.}
{\bf 96}
041303

\end{thebibliography}
\end{document}